\documentclass[epj]{svjour}
\usepackage{graphics}
\usepackage[latin1]{inputenc}
\begin{document}
\hyphenation{Sonder-forschungs-bereich} \hyphenation{Minimum}
\hyphenation{non-ergo-dicity}
\title{Partial clustering prevents global crystallization in a binary 2D colloidal glass former}

\author{F. Ebert\inst{1}, G. Maret\inst{1} \and P. Keim\inst{1}
}                     
%
%
\institute{Fachbereich f\"ur Physik, Universit\"at Konstanz, D-78457
Konstanz, Germany}
\date{Received: \today / Revised version: \today}
%
\abstract{A mixture of two types of super-paramagnetic colloidal
particles with long range dipolar interaction is confined by gravity
to the flat interface of a hanging water droplet. The particles are
observed by video microscopy and the dipolar interaction strength is
controlled via an external magnetic field. The system is a model
system to study the glass transition in 2D, and it exhibits
\textit{partial clustering} of the small particles
\cite{cluster_prl}. This clustering is strongly dependent on the
relative concentration $\xi$ of big and small particles. However,
changing the interaction strength $\Gamma$ reveals that the
clustering does not depend on the interaction strength. The
\textit{partial clustering} scenario is quantified using
\textit{Minkowski} functionals and partial structure factors.
Evidence that partial clustering prevents global crystallization is
discussed.
} 
\maketitle
\section{Influence of dimensionality on frustration}
It is well known that the macroscopic behavior of crystalizing
systems sensitively depends on dimensionality, as demonstrated by
two examples: In 2D an intermediate phase exists between fluid and
crystal, the \textit{hexatic phase}, where the system has no
translational order while the orientational correlation is still
long-range \cite{hexatic_theo,hexatic_exp,hexatic_exp2}. Such a two
step melting scenario is not known in 3D. The \textit{Ising} model
for ferromagnetics shows a phase transition for 2D and 3D but not
for 1D \cite{ising}. For amorphous systems, however, it was found in
experiments \cite{hansroland_glass}, simulations
\cite{harrowell_glass_2d}, and theory \cite{bayer_2d_mct} that the
glass transition phenomenology is very similar in 2D and 3D systems,
both in dynamics and structure \cite{hansroland_glass,epje_ebert}.\\
A subtle difference, the local density optimization in 2D and 3D, is
the following: in 3D the local density optimized structure of four
spheres is obviously a tetrahedron. However, it is not possible to
completely cover space in 3D with tetrahedra, because the angle
between two planes of a tetrahedron is not a submultiple of $360°$
\cite{tetraeder}. The density optimized state with long-range order
is realized by the \textit{hexagonal closed packed} structure or
other variants of the \textit{fcc} stacking with packing fraction
$\phi=\pi/\sqrt{18}\approx74\%$. The dynamical arrest in 3D is
expected to be enhanced by this geometrical frustration, because the
system has to rearrange its local density optimized structure to
reach long-range order\footnote{It is found in 3D hard sphere
systems that this geometrical frustration \textit{alone} is not
sufficient to reach a glassy state as it cannot sufficiently
suppress crystallization \cite{frust1,frust2,frust3}, and
additionally polydispersity is needed \cite{pusey}.}. The local
geometrical frustration scenario is different in 2D. There, the
local density optimized structure and densest long-range ordered
structure are identical, namely hexagonal. For the glass transition
in 2D it is therefore expected that an increase of complexity is
necessary to reach dynamical arrest without crystallization: in
simulations an isotropic one-component 2D system has been observed
undergoing dynamical arrest for an inter-particle potential that
exhibits two length scales, a \textit{Lennard-Jones-Gauss} potential
with two minima \cite{LJG}. Other simulations showed that systems of
identical particles in 2D can vitrify if the mentioned local
geometric frustration is created artificially via an anisotropic
fivefold interaction potential \cite{tanaka_glass_crystal}.
Alternatively, the necessary complexity can be created by
polydispersity as found in simulations \cite{poly_2d}.\\ A
bi-disperse system in general is simple enough to crystallize as
e.g. seen from the rich variety of binary crystal structures in an
oppositely charged 3D coulomb mixture \cite{van_bladeren}. In the
system at hand, \textit{partial clustering} prevents the homogenous
distribution of particles and the system crystallizes locally into
that crystal structure which is closest in relative concentration
\cite{epje_ebert,dipolar_crystals,gen_algorithm}. In this way the
system effectively lowers its energy with a compromise between
minimization of particle transport and minimization of potential
energy. However, that means that the resulting structure is not in
equilibrium, but in a frustrated glassy state. This competition of
local stable crystal structures prevents the relaxation into an
energetically equilibrated state, i.e. a
large mono-crystal.\\
It was found that a binary mixture of magnetic dipoles is a good
model system of a 2D glass former \cite{hansroland_glass} as the
dynamics and structure shows characteristic glassy behavior: when
the interaction strength $\Gamma$ is increased, the system viscosity
increases over several orders of magnitude while the
global structure remains amorphous.\\
For all measured interaction strengths the system shows no
long-range order as probed with bond order correlation functions
\cite{epje_ebert}. However, on a local scale it reveals non-trivial
ordering phenomena:
\textit{partial clustering} and \textit{local crystallinity}.\\
\textit{Partial clustering} \cite{cluster_prl,cluster_pre} means
that the small particles tend to form loose clusters while the big
particles are homogeneously distributed. The heterogeneous
concentration of small and big particles leads to a variety of local
crystal structures when the system is supercooled making up the
globally amorphous structure. This \textit{local crystallinity}
obviously plays a key role for the glass transition in this 2D
colloidal system as it dominates the glassy structure
\cite{epje_ebert}. Therefore, the phenomenon of \textit{partial
clustering} is indirectly responsible for the frustration towards
the glassy state. In section \ref{morph} the details of the
clustering scenario are explained. The dependence on the parameters
accessible in experiment and the relation to \textit{local
crystallinity }\cite{epje_ebert} are discussed in section \ref{mix_cluster}.\\

Firstly, the experimental setup is introduced. After a brief
discussion about origin of \textit{partial clustering}, a
morphological analysis using \textit{Minkowski} measures is
presented to characterize and quantify the effect. Finally, the
dependence on relevant parameters like the average relative
concentration $\xi$ and the interaction strength $\Gamma$ will be
discussed using \textit{Euler} characteristics and the partial
static structure factors.
\section{Experimental setup}
\begin{figure}[h]
\begin{center}
\resizebox{0.75\columnwidth}{!}{\includegraphics{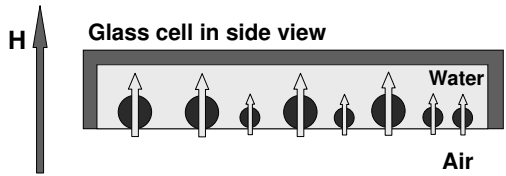}}
\end{center}
\caption[sq] {\label{cell}
  \em Super-paramagnetic colloidal particles confined at a water/air
  interface due to gravity. The curvature of the interface is
  actively controlled to be completely flat; therefore the system
  is considered to be ideal two dimensional.
  A magnetic field $\textbf{H}$ perpendicular to the interface
  induces a magnetic moment $\textbf{m}$ in each bead leading to a repulsive dipolar
  interaction.}
\end{figure}
The detailed experimental setup is explained in \cite{sci_instr}. It
consists of a mixture of two different kinds of spherical and
super-paramagnetic colloidal particles (species A: diameter $d_A=4.5
\:\mu m$, susceptibility $\chi_A=7.4\cdot10^{-11}\:Am^2/T$, density
$\rho_A=1.5\: g/cm^3$ and species B: $d_B=2.8 \:\mu m$,
$\chi_B=6.6\cdot10^{-12}\:Am^2/T$, $\rho_B=1.3\: g/cm^3$) which are
confined by gravity to a water/air interface. This interface is
formed by a water drop suspended by surface tension in a top sealed
cylindrical hole ($6\:mm$ diameter, $1\:mm$ depth) of a glass plate
as sketched in Fig. \ref{cell}. A magnetic field $\vec{H}$ is
applied perpendicular to the water/air interface inducing a magnetic
moment $\vec{M}= \chi \vec{H}$ in each particle leading to a
repulsive dipole-dipole pair-interaction. Counterpart of the
potential energy is thermal energy which generates \textit{Brownian}
motion. Thus the dimensionless interaction strength $\Gamma$ is
defined by the ratio of the potential versus thermal energy:
\begin{eqnarray}\label{glgamma}
\Gamma & = & \frac{E_{magn}}{k_B T} \propto \frac{1}{T_{sys}}\nonumber\\
& = &\frac{\mu_0}{4\pi}\cdot
\frac{\vec{H}^{2}\cdot(\pi\rho)^{3/2}}{k_B
T}(\xi\cdot\chi_B+(1-\xi)\chi_A)^2  \quad .
\end{eqnarray}
Here, $\xi=N_B/(N_A+N_B)$ is the relative concentration of small
species with $N_A$ big and $N_B$ small particles and $\rho$ is the
area density of all particles. The average distance of neighboring
big particles is given by $l_A=1/\sqrt\rho_{A}$. The interaction
strength can be externally controlled by means of the magnetic field
$\textbf{H}$. $\Gamma$ can be interpreted as an inverse temperature
and
controls the behavior of the system.\\
The ensemble of particles is visualized with video microscopy from
below and the signal of a CCD 8-Bit gray-scale camera is analyzed on
a computer. The field of view has a size of $1170 \times 870 \: \mu
m^2$ containing typically $3\cdot10^{3}$ particles, whereas the
whole sample contains about up to $10^{5}$ particles. Standard image
processing is performed to get size, number and positions of the
colloids. A computer controlled syringe driven by a micro-stage
controls the volume of the droplet to get a completely flat surface.
The inclination is controlled actively by micro-stages with a
resolution of $\alpha \approx 1\:\mu$rad. After several weeks of
adjusting and equilibration this provides best equilibrium
conditions for long time stability. Trajectories for all particles
in the field of view can be recorded over several days providing the
whole phase space information.
\section{Origin of partial clustering\label{origin}}
The origin of the clustering phenomenon lies in the negative
\textit{nonadditivity} of the binary dipolar pair potential
\cite{cluster_prl,cluster_pre}. It is not expected in positive
nonadditive mixtures like colloid-polymer mixtures or additive
mixtures like hard spheres. In binary mixtures with additive hard
potentials in 2D, phase separation was found using \textit{Monte
Carlo} simulations \cite{additiv}. In addition to the negative
nonadditivity, the relation $v_{BB}<v_{AB}<v_{AA}$ of the pair
potentials has to be \mbox{fulfilled
\cite{cluster_prl,cluster_pre}}. Why this leads to partial
clustering can be understood as follows: The negative nonadditivity
prevents macro-phase separation as the negativity of the
nonadditivity parameter
$\Delta=2\sigma_{AB}-(\sigma_{AA}+\sigma_{BB})$ means that particles
are effectively smaller in a mixed state
($\sigma_{ij}=\int_{0}^{\infty}dr\{1-\exp[-v_{ij}(r)/k_{B}T]$ are
the \textit{Barker-Henderson} effective hard core diameters). Thus,
a mixed configuration is preferred from this. Additionally, the
inequality $v_{AB}<v_{AA}$ energetically favors direct neighbor
connections between different species instead of big particles being
neighbors. In competition to this, the inequality $v_{BB}<v_{AB}$
favors the neighboring of small particles. The best compromise is
achieved in the partial clustered arrangement: neighboring small
particles that are located
in the voids of big particles.\\
The genuineness of the effect was demonstrated by a comparison of
computer simulation, theory and \mbox{experiment
\cite{cluster_prl}}. There, statistical evidence for the occurrence
of \textit{partial clustering} is provided by the static structure
factor. The structure factor $S_{AA}(k)$ has the characteristic
shape of a one-component fluid. In contrast, the structure factor of
the small particles $S_{BB}(k)$ has a dominant prepeak at small
wave-vectors which is statistical evidence for an inherent length
scale much larger than the typical distance between two neighboring
small particles: it is the length scale of the clusters.\\
The prepeak provides statistical evidence of an inherent length
scale in the small particle configurations. However, the structure
factors do not reveal all details of the phenomenon. For example,
the voids seen in the big particle configurations \mbox{(see Figure
\ref{snapshots_sep})} are not reflected directly in the features of
the partial structure factors. To further elucidate the scenario,
the effect is now investigated from a morphological point of view
using \textit{Minkowski} measures as this provides additional
quantitative insight.
\section{Morphological analysis\label{morph}}
\begin{figure}[t!]
\begin{center}
\resizebox{1.0\columnwidth}{!}{\includegraphics{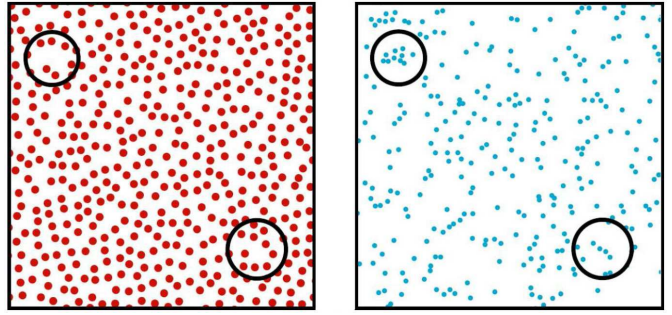}}
\end{center}
\caption[sq] {\label{snapshots_sep} \em Snapshot of a particle
configuration $(\Gamma=5$, relative concentration $\xi=41\%)$ where
big (left) and small (right) particles are displayed separately. The
clusters of small particles (right) fit in the voids formed by the
big particles (left) as highlighted by circles for two examples.}
\end{figure}

For low interaction strengths $\Gamma$ the system is an equilibrated
fluid as seen from the purely diffusive behavior
\cite{hansroland_glass}. Assuming that entropy is maximized it might
be intuitively expected that particles form an arbitrarily mixed
state where small and big particles are evenly distributed. However,
already the inspection of a single snapshot reveals that this is not
the case, and the scenario turns out to be more subtle. How the
system appears in equilibrium at low interaction strengths $\Gamma$
is demonstrated in Figure \ref{snapshots_sep}. There, a
configuration with relative concentration $\xi=41\%$ at $\Gamma=5$
is separately plotted for big (left) and small (right) particles.
Big particles are distributed more evenly while the small particles
form loose clusters. Configurations of big and small particles are
related because small particle clusters are able to push away the
big particles and form voids in the big particle configuration. This
connection becomes obvious in the highlighted regions where two big
clusters of small particles create two voids in the big particle
configuration. This visual impression of the configurational
morphology will be quantified in detail after a brief introduction
of the used tools,
the \textit{Minkowski} measures.\\

\begin{figure}[t]
\begin{center}
\resizebox{0.9\columnwidth}{!}{\includegraphics{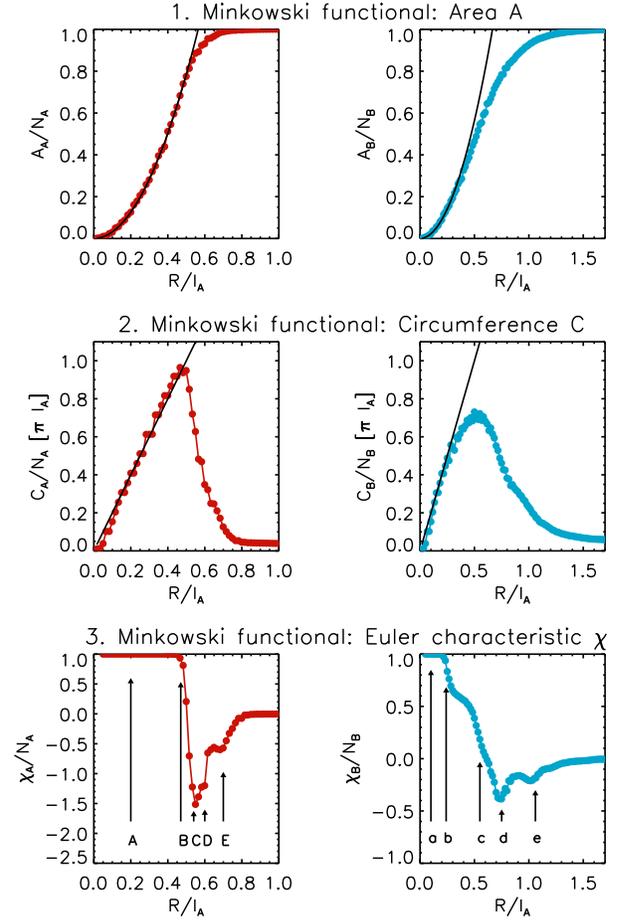}}
\end{center}
\caption[sq] {\label{minkowski_exp} \em The three
\textit{'Minkowski'} measures in 2D ('area', 'circumference', and
'Euler characteristic') are shown for a sample with $\xi=42\%$ at
$\Gamma$=662, separately for big (left graphs) and small particles
(right graphs). Configurations were decorated with discs and the
\textit{'Minkowski'} measures are calculated in dependence on their
radius $R/l_A$. The solid lines correspond to the
\textit{'Minkowski'} measures of a single cover-disc. Marks A-E and
a-e in the bottom graphs correspond to radii of interest, and an
example of a representative configuration at these labels is
displayed in Figure \ref{eulerpics}. The features of all measures
confirm the clustering scenario.}
\end{figure}

\textit{Minkowski} functionals provide morphological measures for
characterization of size, shape, and connectivity of spatial
patterns in $d$ dimensions \cite{hadwiger}. These functionals turned
out to be an appropriate tool to quantify clustering substructures
in \mbox{astronomy}, e.g. from galaxies \cite{euler2}. They also
give insight into the morphology of random interfaces in
microemulsions \cite{euler1}.\\The scalar \textit{Minkowski}
valuations $V$ applied to patterns $P$ and $Q$ in \textit{Euclidian}
space are defined by three types of covariances \cite{euler2}:
\begin{enumerate}
  \item Invariance to motion;
  \item Additivity: $V(P \bigcup Q)= V(P)+V(Q)-V(P \bigcap Q)$;
  \item Continuity: continuous change for slight distortions in pattern P.
\end{enumerate}
It is guaranteed by the theorem of \textit{Hadwiger} that in $d$
dimensions there are exactly $d+1$ morphological measures $V$ that
are linearly independent \cite{hadwiger}. For $d=2$, the three
functionals have intuitive correspondences\footnote{For d=3 a common
set of functionals correspond to: volume, area, integral mean
curvature, and \textit{Euler characteristic}.}: The
\textbf{\mbox{surface area}}, the \textbf{circumference} of the
surface area, and the \textbf{Euler
characteristic $\chi$}.\\
In two dimensions the \textit{Euler characteristic} $\chi$ for a
pattern $P$ is defined as
\begin{equation}\label{eulerchardefinition}
  \chi=S-H
\end{equation}
where $S$ is the number of connected areas and $H$ the number of holes.\\
Morphological information can also be obtained from particle
configurations. As configurations only consist of a set of
coordinates, a cover disc with radius $R$ is placed on each
coordinate to construct
a pattern that can be evaluated.\\
The \textit{Minkowski} measures are then determined for different
cover radii $R$, leading to a characteristic curve for a given
configuration, explained in the following (for better understanding follow curves in Figure \ref{minkowski_exp})\\
The first \textit{Minkowski} measure (disc area normalized to total
area) increases from $0\%$ to $100\%$ for increasing radius $R$ with
a decreasing slope when discs start to overlap. The second
\textit{Minkowski} measure (circumference) increases with cover
radius $R$, reaches a maximum, and then decays to zero when all
holes are overlapped. The third \textit{Minkowski} measure
(\textit{Euler characteristic}) is very subtle and describes the
connection of cover discs and the formation of holes. It allows
the most detailed interpretation of a given configuration.\\
A typical devolution of an \textit{Euler characteristic} $\chi/N$
with N particles can be divided in three characteristic parts for
continuously increasing cover disc radius $R$:
\begin{enumerate}
  \item For small $R$ the curve is constant at $\chi/N=1$ (normalized to the number of particles $N$). Discs are not touching and thus $S$ is equal to the number of
  particles. No holes are present. As $\chi$ is normalized with the number
  of points, the \textit{Euler characteristic} amounts to $\chi/N=1$.
  \item With increasing $R$ the curve drops and $\chi/N$ can become negative when cover discs are
  large enough to overlap and holes are formed. Therefore, the number of connected areas $S$ decreases and
  holes are forming which further decreases $\chi/N$. The minimum is
  reached when discs are connected to a percolating network and the maximum number of holes has formed.
  \item For large $R$ the curve starts to raise again because the holes are collapsing until
  the whole plane is covered with overlapping discs and $\chi/N \rightarrow 0$ for $R \rightarrow
  \infty$.
\end{enumerate}

\begin{figure*}[t]
\begin{center}
\resizebox{1.8\columnwidth}{!}{\includegraphics{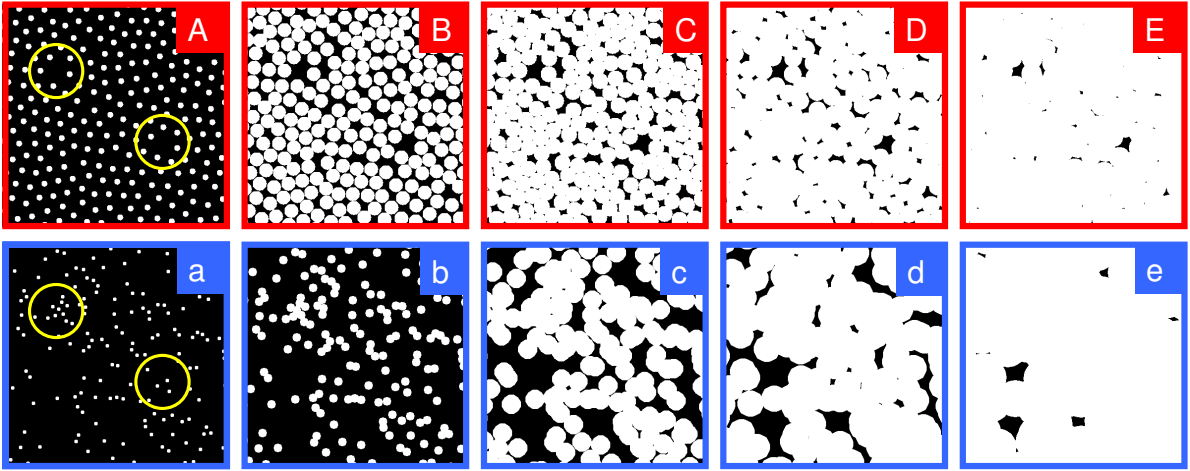}}
\end{center}
\caption[sq] {\label{eulerpics} \em The same section
($375\times375\,\mu m^2$) of a single particle configuration with
cover-discs is shown several times in the pictographs A-E and a-e
with increasing radius. Pictures correspond to certain radii of
interest marked by arrows in the corresponding graphs of the 'Euler
characteristic' in Figure \ref{minkowski_exp}. In the first (second)
row, just the big (small) particles of this particular configuration
are shown. Voids in the big particle configurations are filled with
clusters of small particles (examples are highlighted with circles
in pictograph 'A' and 'a' respectively). Configurations were
obtained from a sample with $\xi=42\%$ at $\Gamma=662$.}
\end{figure*}
This qualitative behavior is typical for configurations in 2D.
However, the individual morphological information is obtained from
specific features in the three regions as the onset of the fall and
raise, characteristic kinks or plateaus, and the slope of the fall
and raise. The \textit{Minkowski} measures in dependence on an
increasing cover disc provide a characteristic morphological
'fingerprint' of configurations and therefore statistical evidence
of the clustering scenario, complementary to structure factors
\cite{cluster_prl,cluster_pre}. The statistical noise of the curves
is remarkably small compared to that of structure factors as the
whole statistical information of a configuration is contained in
\textit{every} data point. Thus, even small features in the curves
are true evidence for morphological particularities.\\
All three \textit{Minkowski} measures in
2D, \textit{area}, \textit{circumference}, and \textit{Euler
characteristic}, are averaged over $100$ configurations for a given temperature and no time-dependence was found during this period of about half an hour. The curves are
plotted separately for both species in Figure
\ref{minkowski_exp} in dependence on the cover-disc radius.
Corresponding snapshots are displayed in Figure \ref{eulerpics} to
illustrate characteristic radii as indicated. The used sample was
strongly supercooled ($\Gamma=662$, $\xi=42\%$), i.e. it was not in
equilibrium. However, the features found at these high interaction
strengths are the same as for low $\Gamma$, where the sample is in
equilibrium. The high interaction strength is used here, as the
discussed features become clearer, but it is assumed in the
following that the conclusions on clustering are also valid for low
interaction \mbox{strengths $\Gamma$.} This assumption will be
justified when the dependency on
interaction strength is discussed in section \ref{mix_cluster}.\\
\textbf{First Minkowski measure: Area}\\ The upper graphs of Figure
\ref{minkowski_exp} show the area per particle in dependence on the
cover-radius $R/l_A$. The covered area starts at zero and is
increased continuously to $100\%$ when the discs completely overlap
the area. The solid line in both plots indicates the area fraction
covered by non-touching free discs. The deviation of the first
\textit{Minkowski} measure from that line shows how homogeneous a
configuration is. A clear difference is found between particle
species: The big particles follow this reference line up to
$\approx80\%$. This is close to the maximum possible value for hard
discs between $84\%$ and $90.7\%$ for \textit{random close packing}
and \textit{hexagonal close packing} respectively
\cite{bayer_2d_mct,random_close,random_close2}. Therefore, the big
particle configurations are very homogeneously distributed. However,
the small particles curve deviates from the free disc reference much
earlier at $\approx40\%$ indicating that small particles are much
less evenly spread, i.e. they form clusters.\\
\textbf{Second Minkowski measure: Circumference}\\ The middle graphs
show how the circumference depends on the radius. The black
reference line corresponds to a free expanding circle. The measure
of the big particles follows this reference line up to
$R/l_A\approx0.5$ and then sharply decreases. Again, this is due to
the homogeneous distribution of the big particles. Their cover-discs
can expand freely up to the maximum possible value of
$R_{max}/l_A=1/2$ for square order. The measure of the small
particles deviates much earlier and less steeply as their particle
density is very heterogeneous, i.e. clustered.\\
\textbf{Third Minkowski measure: Euler Characteristic}\\ The most
detailed information on morphology is obtained from the
\textit{Euler characteristic}. It exhibits many features related to
characteristic structures of the investigated configurations. For
better understanding of the features described in the following,
Figure \ref{eulerpics} shows snapshots of a typical section in the
used configuration for specific cover-radii. The notation $A-E$ for
the big particles and $a-e$ for the small ones is used in both
Figures. In Figure \ref{eulerpics}a and \ref{eulerpics}A two
clusters are highlighted. Firstly, we consider only the big
particles. The \textit{Euler characteristic} $\chi_A$ per particle
is 1 as the expanding discs are not touching for small radii (mark
A). Again, this continues up to a value close to $R/l_A\approx0.5$.
The characteristic deviation of all three measures at this same
radius states the homogeneity of the big particle distribution.
Then, discs touch and $\chi_A/N_A$ decreases rapidly because
surfaces are connecting and holes are forming (mark B). The minimum
is reached at \mbox{mark C}, and the \textit{Euler characteristic}
immediately raises because the smallest holes between the triangular
close packed regions collapse as seen in the comparison of Figures
\ref{eulerpics}C and \ref{eulerpics}D. The next holes to collapse
are those where one small isolated particle is located. Therefore, a
little kink is visible at mark D since these one-particle holes are
a little larger than the holes decaying at mark C and therefore
'survive' a little longer. When they collapse, the \textit{Euler
characteristic} increases rapidly to a pronounced plateau. Note,
this plateau is the only statistical evidence for the voids in the
big particle configuration made up by the clusters: these voids are
large and thus they 'survive' for a long 'time' resulting in that
plateau. These voids are not detectable with the other
\textit{Minkowski} measures or the static structure factor of the
big particles \cite{cluster_prl,cluster_pre}. Finally, they start to
decay at mark E, but not
suddenly, which shows that they have a distribution in size.\\
The \textit{Euler characteristic} of the small particles shows the
complementary picture: Starting with low values of $R/l_A$ the
characteristic is $\chi_B/N_B=1$ for free disc expansion. The first
drop at mark b occurs at much lower values than for the big
particles because small particles in clusters connect. The
subsequent shoulder right next to mark b confirms the clustering:
small particles inside a cluster are now connected, and it needs
some further increase of disc radius until the clusters themselves
start connecting. A small second shoulder at mark c originates from
the isolated particles that are not arranged in clusters. They are
the last particles incorporated until all discs form a percolating
network at the minimum at \mbox{mark d}. The increase of
$\chi_B/N_B$ shows how the holes are closing. While the increase in
the \textit{Euler characteristics} of the big particles has a
plateau at \mbox{mark E}, the small particles have a clear dip at
mark e. This reveals information on the shape of the clusters: The
voids in the big particle positions are compact in shape stopping
the increase of the \textit{Euler characteristic} before mark E. In
contrast, the small particles arrange in chain-like clusters. When
the voids between these structures close, they decay into several
sub-holes causing the characteristic to decrease again. In fact, the
big particles also cause a little dip at their plateau for the voids
can sometimes also decay into sub-holes. However, this dip is much
smaller than for the small ones.\\ Most features are also visible in
the \textit{Euler characteristic} obtained from \textit{Brownian
dynamics} simulation \cite{cluster_prl}. There, the same qualitative
behavior is found but the smaller features are 'washed out' because
the used interaction strengths were much lower, as discussed in the
following section. Further, the variation of the relative
concentration $\xi$ and the subsequent dependence of the features in
the \textit{Euler characteristic} will confirm the interpretation of
the scenario.

\section{\label{mix_cluster}Dependence of clusters on interaction strength and relative concentration}
\begin{figure}[t]
\begin{center}
\resizebox{0.9\columnwidth}{!}{\includegraphics{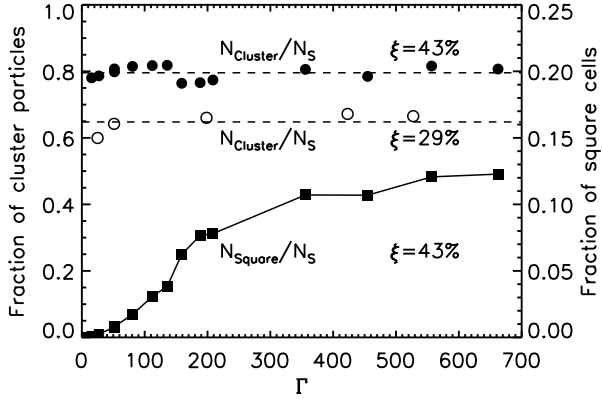}}
\end{center}
\caption[sq] {\label{clusterfraction} \em Fraction of small
particles (relative to all small particles $N_S$, left axis) that
are arranged in clusters. This fraction is independent of $\Gamma$
but dependent on relative concentration $\xi$. The dashed lines
correspond to the
average fractions over all interaction strengths $\Gamma$.\\
Solid squares represent the fraction of small particles (relative to
all small particles $N_S$, right axis) that are arranged in square
symmetry as evaluated in \cite{epje_ebert}.}
\end{figure}
In order to demonstrate the connection between clustered
equilibrated fluid and supercooled local crystalline structure, the
dependence of partial clustering on the interaction strength
$\Gamma$ and on the
average relative concentration $\xi$ is now discussed.\\
In Figure \ref{clusterfraction} the fraction of small particles
arranged in clusters is plotted versus interaction parameter
$\Gamma$ for two different relative concentrations $\xi$. A small
particle is characterized as 'cluster-particle', when the closest
neighbor is also a small particle. This simple criteria implies that
the smallest possible cluster consists of two close small particles
surrounded in a cage of big ones. In the graph of Figure
\ref{clusterfraction} for $\xi\approx 43\%$ it is found that a high
fraction of $\approx 80\%$ of all small particles is arranged in
clusters. Even for a lower relative concentration $\xi\approx 29\%$,
still $\approx 65\%$ are arranged in clusters. Note, that the
fraction of small particles in both samples is smaller than that of
the big particles as $\xi<0.5$. Therefore, every small particle
could have enough possibilities to arrange far away from the next
small particle which is obviously not the case. For an arbitrary
distribution of the small particles over the number of possible
sites (which is equal to the number of big particles) a fraction of
$40\%$ is expected for a relative concentration $\xi=29\%$ and a
fraction of $55\%$ for a relative concentration $\xi=43\%$
\footnote{A simple simulation is performed where $N_A$ sites are
randomly occupied with small particles, and the same analysis to
determine the number of cluster-particles is applied.}. The fact
that these expected values are significantly lower than the actually
measured ones additionally confirms that small particles
effectively attract each other and therefore cluster.\\

The main result from Figure \ref{clusterfraction} for the structure
of this colloidal glass former is that the fraction of
cluster-particles is independent of the interaction parameter
$\Gamma$, in contrast to local crystallinity which strongly
increases upon supercooling: Clusters do not vanish although the
local structure is dominated by local crystallinity for strong
supercooling \cite{epje_ebert}. This is demonstrated for the example
of square order in the same Figure
\ref{clusterfraction} (for details see \cite{epje_ebert}).\\
The local relative \textit{concentration} is frozen in. Small
particles are not redistributed to match an equilibrium crystal
structure which would reduce the number of cluster-particles (e.g.
in square order). In fact, the independence of the clustering from
$\Gamma$ shows that the opposite is the case: The clusters force the
local structure into that crystalline order which matches best with
the local relative concentration. In this way, local crystallinity
is established without long-range order \cite{epje_ebert} as it
inherits the clustered distribution of the small particles.\\
\begin{figure}[t]
\begin{center}
\resizebox{1.0\columnwidth}{!}{\includegraphics{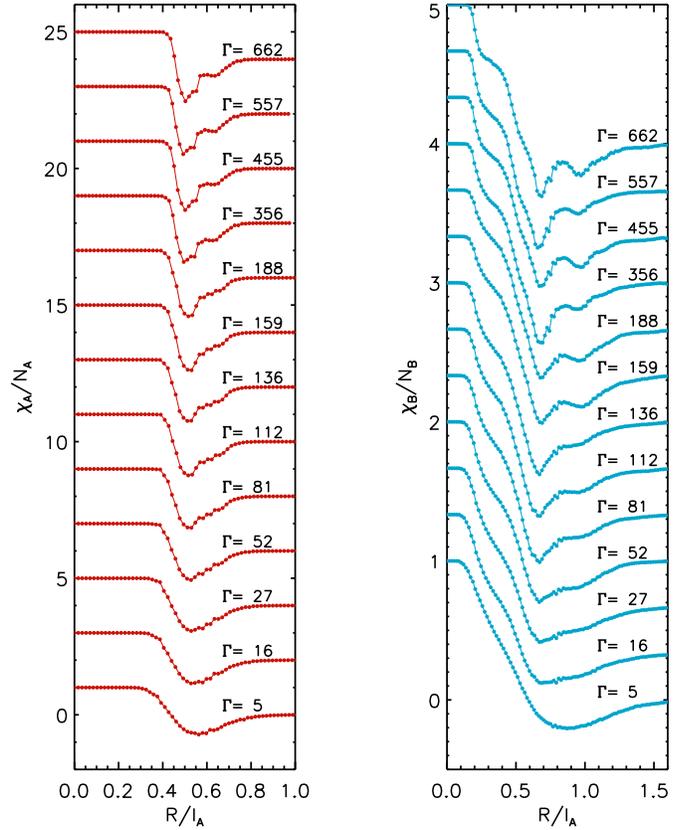}}
\end{center}
\caption[sq] {\label{euler_all} \em 'Euler characteristics' of big
(left) and small (right) particles for different interaction
strengths $\Gamma$. Characteristic features are visible for the
lowest $\Gamma$ and become clearer with increasing $\Gamma$. The
relative concentration was $\xi\approx43\%$. Curves are shifted for
reasons of clarity.}
\end{figure}
This behavior is confirmed by the graphs of Figure \ref{euler_all}.
The \textit{Euler \mbox{characteristics}} for both species are
plotted over a wide range of the interaction \mbox{strength
$\Gamma$}, from fluid $(\Gamma=5$) to the strongly supercooled state
$(\Gamma=662$): the curves change continuously. The main clustering
features as discussed in Figure \ref{morph} are visible for all
values \mbox{of $\Gamma$}, they just become sharper with increasing
interaction strength. The smallest features like the kink at mark D
in Figure \ref{minkowski_exp} are smeared out for low $\Gamma$ but
the plateaus and shoulders characterizing the \textit{partial
clustering} are qualitatively independent of
$\Gamma$.\\

\begin{figure} [tb]
\begin{center}
\resizebox{1.0\columnwidth}{!}{\includegraphics{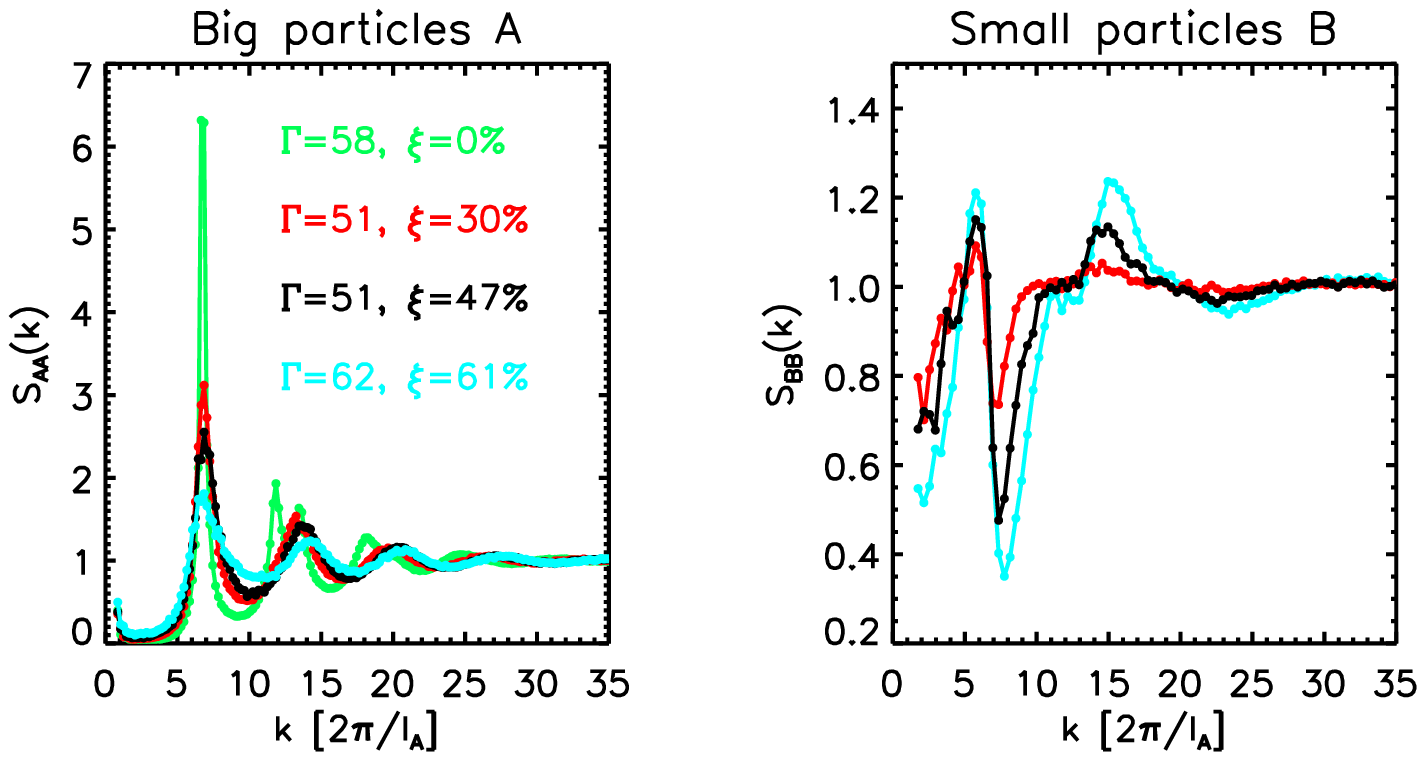}}
\resizebox{1.0\columnwidth}{!}{\includegraphics{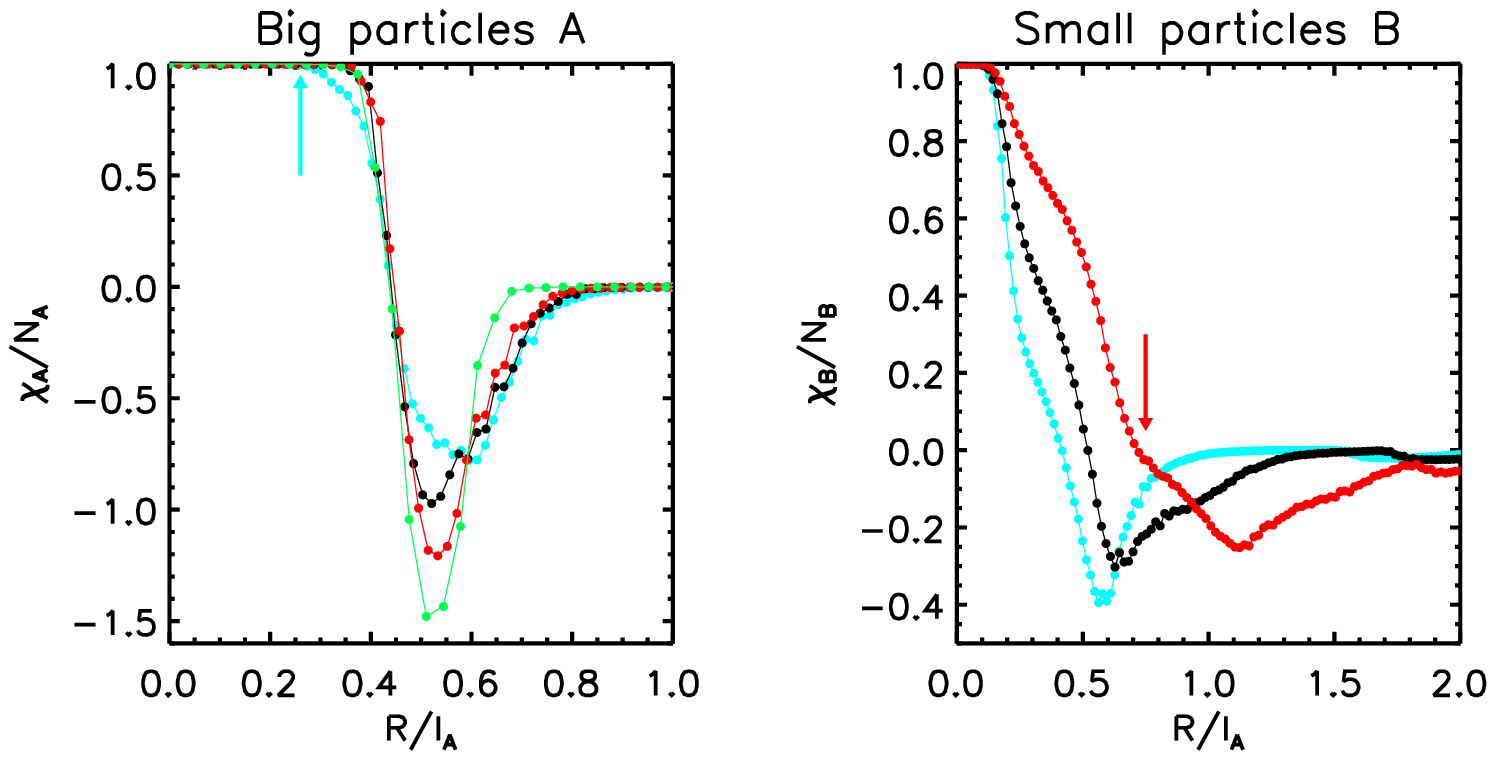}}
\end{center}
\caption[temob] {\label{peak_shift} \em Static structure factors and
\textit{'Euler characteristics'} of samples with comparable
interaction \mbox{strength $\Gamma$} but different relative
concentration $\xi$. Indicated values \mbox{for $\Gamma$} and $\xi$
are valid for all graphs.\\ Upper graphs: Static structure factors
for big (left) and small particles (right). With increasing relative
concentration of small particles the features of $S_{BB}(k)$ gain
more contrast, and peaks in the $S_{AA}(k)$ are shifted slightly
towards higher
k-values.\\
Lower graphs: \textit{'Euler characteristics'} for big (left) and
small particles (right). For decreasing $\xi$ the minimum in
$\chi_{B}/N_B$ is shifted towards larger radii $R$. The weight of
the features changes, e.g. the last shoulder in the drop in
$\chi_{B}/N_B$ (red arrow, right plot). This shoulder results from
the incorporation of isolated small particles into the network (see
Figure \ref{minkowski_exp}) and is only visible for the lowest
relative concentration $\xi=30\%$ (red curve) at this interaction
\mbox{strength $\Gamma$}. The blue arrow in the bottom left plot
marks the onset of the inter-particle connection of the big
particles for the sample with highest relative concentration
$\xi=61\%$. The onset is shifted to a lower value $R$ compared to
the other curves.}
\end{figure}
In Figure \ref{peak_shift} the dependence of the local structure on
relative concentration $\xi$ is shown using structure factors and
\textit{Euler characteristics}. There, samples with comparable
interaction strengths $\Gamma$ but different relative concentrations
$\xi$ are compared. Adding small particles is shifting the peaks of
the structure factor $S_{AA}(k)$ towards higher $k$-values. This can
be understood by the clustering effect: Small particles form
clusters and push the big particles closer together resulting in a
shift of the main peak. This shift is small for the used parameters.
However, confirmation of this interpretation is found in
\cite{cluster_pre} where \textit{Liquid integral equation theory}
shows the same result unambiguously. There, parameters were used
that are not accessible in the experiment (different ratios of the
magnetic moments $\chi_B/\chi_A$). The contrast in $S_{BB}(k)$ is
increased for higher relative concentrations $\xi$ which is also
in agreement with \mbox{theory \cite{cluster_pre}}.\\
The \textit{Euler characteristics} for the same samples, shown in
the lower graphs \mbox{of Figure \ref{peak_shift}}, confirm this
interpretation. The drop in the $\chi_{A}/N_A$ (bottom left) becomes
deeper when less small particles are present. Then, the distances
between big particles are less distributed due to fewer clusters.
The increase becomes steeper for the same reason: clusters of small
particles cause larger voids collapsing at higher cover-disc radii.
It is remarkable, that the onset of the steep drop is earlier for
high relative concentrations ($\xi=61\%$, blue curve) as indicated
by the blue arrow. Again, this is caused by the small particle
clusters that push together the big particles.\\ A strong dependence
is found in the \textit{Euler characteristics} of the small
particles: the onset of the first drop is independent on the
relative concentration indicating that the local density of
particles in clusters is not affected (unlike that of the big ones).
What significantly changes is the depth of the first drop. The more
small particles, the deeper the drop, because more small particles
are arranged in clusters. The last shoulder, before the
\textit{Euler characteristic} reaches its minimum (marked by red
arrow), refers to the isolated particles (see section \ref{morph}).
Therefore, at these interaction strengths this shoulder is only
visible for the sample with the lowest relative concentration
$\xi=30\%$ (red curve) which has the most isolated particles
(compare also with Figure
\ref{clusterfraction}).\\
The systematic dependence of \textit{Euler characteristics} and
static structure factors on relative concentration $\xi$ confirms
the interpretation of \textit{partial clustering} of section
\ref{morph}. However, the main result of this section is that the
principle occurrence of the effect is independent of the interaction
strength: The clustering in equilibrium at low interaction strengths
is therefore responsible for the variety of local crystallinity at
strong supercooling suppressing long-range order \cite{epje_ebert}.

\section{Conclusions}
On a local scale the system reveals the non-trivial ordering
phenomenon of \textit{partial clustering}: the small particles tend
to form loose clusters while the big particles are homogeneously
distributed. The origin of this effect is traced back to the
negative \textit{nonadditivity} of the dipolar pair potential. The
detailed scenario is quantified using \textit{Minkowski} functionals
applied to experimentally obtained configurations. Changing the
interaction strength $\Gamma$ reveals that the principle scenario
does not qualitatively depend on the interaction strength, and, as a
consequence, the local relative concentration is simply 'frozen' in.
However, the strength of the effect increases with the relative concentration $\xi$.\\
The clustering effect together with the missing ability of the
system to reorganize fast enough into an equilibrated state (i.e.
extended crystal structure
\cite{epje_ebert,dipolar_crystals,gen_algorithm}) is crucial to
understand the glass forming behavior of this system: The
\textit{partial clustering} leads locally to a heterogeneous
relative concentration $\xi$ which then leads for increasing
interaction strengths $\Gamma$ to \textit{local crystallinity}
\cite{epje_ebert} without long-range order. It provides the
necessary complexity for glassy frustration in this 2D system and
prevents solidification into the energetically preferred crystalline
or poly-crystalline
morphologies \cite{dipolar_crystals,gen_algorithm}.\\

This work was supported by the Deutsche Forschungsgemeinschaft SFB
513 project B6, SFB TR 6 project C2 and C4 and the International
Research and Training Group "Soft Condensed Matter of Model Systems"
project A7. We thank P. Dillmann for fruitful discussion and
experimental contributions.

\end{document}